\documentclass[
    a4paper,prb,
    bibnotes,
    twocolumn,
    showpacs,
    superscriptaddress,
    groupedaddress,
]{revtex4}

\usepackage{graphicx}
\usepackage{amssymb,amsmath}
\usepackage{amsmath}
\usepackage{fullpage}

\newcommand{\beq}{\begin{equation}}
\newcommand{\eeq}{\end{equation}}
\newcommand{\bra}[1]{\langle #1 |}
\newcommand{\ket}[1]{| #1 \rangle}
\newcommand{\brahket}[3]{\langle #1 | #2 | #3 \rangle}
\newcommand{\braket}[2]{\langle #1 | #2 \rangle}
\newcommand{\ketbra}[2]{\ket{#1}\bra{#2}}

\newcommand{\mysum}[2]{\sum\limits_{#1}^{#2}}
\newcommand{\myint}[2]{\int\limits_{#1}^{#2}}

\newcommand{\mybpar}[1]{\left( #1 \right)}
\newcommand{\mymat}[2]{\left[ {\bf #1} \right]_{#2}}
\newcommand{\beqa}{\begin{eqnarray}}
\newcommand{\eeqa}{\end{eqnarray}}

\newcommand{\etal}{\mbox{\textit{et al.}}~}

\newcommand{\Figref}[1]{Fig.~\ref{#1}}

\newcommand{\Appref}[1]{Appendix \ref{#1}}

\begin{document}
\title{Transmission eigenchannels from non-equilibrium Green's functions}
\date{\today}


\author{Magnus \surname{Paulsson}}
\email{mpn@mic.dtu.dk}
\affiliation{Division of Physics, School of Pure and Applied Natural Sciences,
University of Kalmar, 391 82 Kalmar, Sweden}
\affiliation{Department of Micro- and Nanotechnology (MIC), Technical
    University of Denmark (DTU), {\O}rsteds Plads, Bldg.~345E, DK-2800
    Lyngby, Denmark}
\affiliation{Department of Electronics, Toyama University, Gofuku,
Toyama,  930-8555, Japan}
\author{Mads \surname{Brandbyge}}
\email{mbr@mic.dtu.dk}
\affiliation{Department of Micro- and Nanotechnology (MIC), Technical
    University of Denmark (DTU), {\O}rsteds Plads, Bldg.~345E, DK-2800
    Lyngby, Denmark}

\begin{abstract}
The concept of transmission eigenchannels is described in a
tight-binding nonequilibrium Green's function (NEGF) framework. A
simple procedure for calculating the eigenchannels is derived using
only the properties of the device subspace and quantities normally
available in a NEGF calculation. The method is exemplified by visualization
in real-space of the eigenchannels for three different molecular and 
atomic-wires. 
\end{abstract}


\maketitle


\section{Introduction}

Electronic transport properties of atomic-scale conductors has been
investigated intensively in the last
decade\cite{AgYeva.03.Quantumpropertiesof,Cuniberti2005}. Examples
of interest include molecular wires connected to metal electrodes,
atomic metal wires, and nanotubes. First principles transport
calculations on these systems give results that are, in general,
difficult to interpret due to the multi-channel nature of the
scattering problem and the fact that the scattering states are
generated from the atomic valence orbitals. The free-electron-type
of reasoning normally used in mesoscopic quantum transport is thus
not adequate. It is therefore useful to analyze the conduction in
terms of eigenchannels. Eigenchannels are particular scattering
states \cite{B.88.Coherentandsequential} with a well-defined
transmission probability, $0\le T_n\le 1$, where the individual
eigenchannel transmissions add up to the total transmission
$T=\sum_n T_n$. In addition to being useful for analyzing
theoretical calculations, the eigenchannel transmissions may be
obtained experimentally (i) with superconducting electrodes
connecting the atomic-scale conductor, as first shown by Scheer
\etal \cite{ScAgCu.98.signatureofchemical}, or (ii) from shot noise
measurements\cite{AgYeva.03.Quantumpropertiesof} where information
about the individual channel contributions can be obtained since the
Fano factor involve the sum\cite{AgYeva.03.Quantumpropertiesof,
Buttiker.Noise} $\sum_n T_n (1-T_n)/\sum_n T_n$.

Eigenchannels have previously been calculated for atomic metal
wires\cite{KoBrTs.00.First-principlesstudyof,SamoriBook} and
molecular contacts\cite{TaBrSt.03.Conductanceswitchingin} by
directly solving for the scattering states in the leads. This type
of analysis breaks up the transmission into the 'non-mixing'
channels\cite{B.88.Coherentandsequential,
BrSJa.97.Conductanceeigenchannelsin, CuYeMa.98.MicroscopicOriginof}
and gives an intuitive picture of electron transport. The ability to
plot the eigenchannel wavefunctions is especially useful, since it
gives a direct spatially resolved picture of the orbitals involved
in the transport. Another possibility is to consider projections of
eigenchannel wavefunctions onto, for example, molecular orbitals.

An increasingly popular theoretical approach to calculate
transport properties is the non-equilibrium Green's function (NEGF)
formalism\cite{Datta}. This is normally used in combination with a
tight-binding type or LCAO electronic structure description\cite{
BrMoOr.02.Density-functionalmethodnonequilibrium,Damle.CP02,Rocha.PRB06,
Palacios.PRB02}.
In this approach, it is straightforward to calculate the single
particle (retarded) Green's function (matrix) including coupling
to the infinite electrodes by introducing self-energies.
The Green's
function is thus the fundamental quantity in these calculations and
scattering states are normally not considered.
Interpretation of the results in terms of scattering states is therefore
non-trivial\cite{Palacios.PRB06}.
In contrast, the
scattering states are the fundamental quantity in approaches based
on the Lippmann-Schwinger equation where jellium models are normally
used to describe the electrodes\cite{LA.95.RESISTANCEOFATOMIC,HiTs.95.,
Paulsson.PRB2001,DiLa.02.Transportinnanoscale}.

\begin{figure}[tbh!]
\begin{center}
\includegraphics[width=0.85 \columnwidth,angle=0]{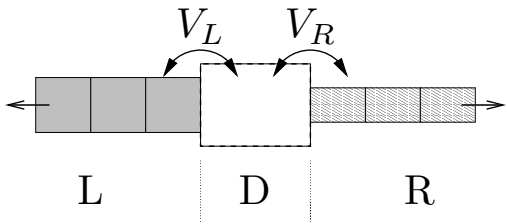}
\end{center}
\caption{The generic two probe system which
couples the left ($L$) and right ($R$) leads through an
intermediate "device" region ($D$). \label{fig.sys}}
\end{figure}

The aim of this paper is to show how the eigenchannels can be easily
generated within the NEGF approach {\it without} solving for the
scattering states in the leads. The eigenchannel wavefunctions are
here obtained directly from quantities readily available in the NEGF
calculation, e.g., the retarded Green's function matrix, $G_D$ of the
device region, and the $\Gamma_{L,R}$ matrices describing the
coupling of the device region to the two electrodes ('left' $L$ and
'right' $R$, see Fig.~\ref{fig.sys}). In the case of atomistically
defined electrodes, this approach is especially advantageous since
solving for the scattering states requires calculating the Bloch
waves in the electrodes (complex band structure), which may be a
non-trivial numerical task for large unitcells\cite{Sankey.PRB02}.
Related to our approach is the so-called (left/right) open 'channel
functions' of Inglesfield an co-workers based on the 'embedding
potential' in the real-space
formulation\cite{InCrIs.05.Embeddingpotentialdefinition}, and the
KKR-based formulation by Bagrets \etal \cite{Bagrets,Bagrets2006}.
In addition to providing a simple way to calculate the
eigenchannels, the method presented here offers an intuitive
understanding of the one-particle NEGF equations, e.g., it may be
used to understand propensity rules for the effect of phonon
scattering on the electronic
transport\cite{Paulsson.PRB05,Paulsson.unpub07}.

The paper is organized as follows. Sec.~\ref{sec.2} starts with the
definition of eigenchannels and a summary of the standard one
particle NEGF equations. Our method to obtain the eigenchannels
without solving for the full transmission matrix is then derived.
The usefulness of the eigenchannels is illustrated in
Sec.~\ref{sec.3} with three examples where the eigenchannels are
calculated for molecular and atomic-wires connected to gold
electrodes using a first-principles density functional method.

\section{Eigenchannels}
\label{sec.2}

We consider transport through a device region, $D$, coupled to two
semi-infinite leads, left and right ($L, R$) and limit ourselves to
leads built from periodic cells. The solutions in the
corresponding infinite left/right leads are Bloch states
$\ket{u_{l}}$ ($\ket{u_{r}}$) denoted by band-index $l$ ($r$) for the
left (right) lead. In order to obtain the transmission amplitude matrix
${\bf t}_{r,l}$ at a given energy, $E$, we consider the solutions to the
Schr\"odinger equation, $\ket{\Psi_l}$ with scattering boundary
conditions and incoming waves in the left lead,
\beqa
  \ket{\Psi_l}=&
    \frac{\ket{u_l}}{\sqrt{v_l}} + \text{Reflected}
       & \mbox{; left lead}, \nonumber \\
  \ket{\Psi_l}=&
    \mysum{r}{}\frac{\ket{u_r}}{\sqrt{v_r}}{\bf t}_{r,l} + \text{Decaying}
       & \mbox{; right lead} , \label{eq.scatbound}
\eeqa where $\ket{u_l}$ is an incoming Bloch wave from the left
lead, i.e., right-moving, with energy $E$ and group velocity
$v_l>0$, and $\ket{u_r}$ is an out-going Bloch wave ($v_r>0$). At a
certain energy, the number of such in-coming channels, $N_L(E)$, is
determined by the band structure of the lead. Likewise, there are
$N_R(E)$ such out-going channels on the right. In
Eq.~\ref{eq.scatbound}, we explicitly state the {\em
flux-normalization} by dividing the Bloch state by ${\sqrt{v_l}}$,
where the Bloch waves $\ket{u_l}$ are  normalized in the
conventional manner over the infinite leads\footnote{The flux $v_l$
is the band-velocity divided by the $L$ unit-cell length, while the
left lead $k$-vector is in units of the inverse unit-cell lengths.
Likewise for $R$. Alternatively, $k$ has a unit of inverse length
while the $v$'s are in units of velocity. In any case the unit of
the flux normalized wavefunctions is square root of time.}
$\braket{u_l(k)}{u_{l'}(k')}=\delta_{ll'}\,\delta(k-k')$ . The
reflected and transmitted parts may contain evanescent decaying
waves which have zero velocity and for those states we use normal
integral normalization. With these considerations we find that the
flux-normalized states fulfil the normalization, \beq
\frac{1}{\sqrt{ v_l}}\braket{u_l(k)}{u_{l'}(k')} \frac{1}{\sqrt{
v_{l'}}}=\, \hbar \, \delta_{ll'}\,\delta(E-E') \, , \eeq since the
velocity is related to the energy as $v_l=1/\hbar\,
d\varepsilon_l(k)/dk$. The scattering states generated from the
flux-normalized incoming waves will also be flux-normalized, see
\Appref{app.orthogonal}, \beq \braket{\Psi_l}{\Psi_{l'}}=\hbar \,
\delta_{ll'} \, \delta(E-E') \, , \eeq and likewise for $r,r'$ while
$\braket{\Psi_l}{\Psi_{r}}=0$.

The advantage of using flux-normalized states is that we can make
any unitary transformation between the incoming scattering states
from the left lead at a particular energy, \beq \ket{w_l}= \sum_{l'}
\frac{\ket{u_{l'}}}{\sqrt{v_l}} \, ({\bf U}_L)_{l',l}
\label{eq.lchan} \, . \eeq These new states will again solve the
Schr\"odinger equation, only contain incoming waves in the left
lead, and be fluxnormalized. However, the mix of Bloch waves will no
longer have Bloch symmetry. Naturally we can apply a similar
transformation of the out-going right channels with ${\bf U}_R$.
Especially we can choose the transformations $({\bf U}_L,{\bf U}_R)$
such that the transmission matrix ${\bf t}$ becomes a diagonal
matrix\cite{BrSJa.97.Conductanceeigenchannelsin}, ${\bf t}_e$, at a
specific energy
\beq {\bf t}_e = {\bf U}_R^\dagger \,{\bf t} \,{\bf
U}_L = {\mbox{Diag}}\left(\sqrt{T_1},\sqrt{T_2},\hdots \right) .
\label{eq.tdiagA}
\eeq
This corresponds to a singular value
decomposition (SVD) of the transmission amplitude matrix, or a
diagonalization of the Hermitian (left-to-right) transmission
probability matrix,
\begin{equation}\label{eq.tdiagT}
{\bf T}_{e}={\bf t}_e^{\dagger}{\bf t}_e = {\bf U}_L^\dagger
\,{\bf t}^\dagger{\bf t} \,{\bf U}_L = {\mbox{Diag}}\left(T_1,
T_2 ,\hdots \right) \, ,
\end{equation}
where there will be a maximum of ${\mbox{Min}}(N_L(E),N_R(E))$
non-zero, eigenvalues $0 \le T_\alpha \le 1$.

The diagonalization of the transmission matrix defines the
transmission {\em eigenchannels} ${\ket{\Phi_\alpha}}$ of the system
as the unitary mix of left in-coming flux-normalized scattering states
(``channels'') given by
${\bf U}_{L}$, i.e., the left eigenchannels are given by
${\ket{\Phi_\alpha}} = \mysum{l'}{}{\ket{\Psi_{l'}} ({\bf
U}_L)_{l',\alpha}}$. The eigenchannels have the special property of
being ``non-mixing'' in the sense that the transmission of a sum of two
is the sum of individual fluxes or transmissions (equal for
flux normalized states). For
example, consider the alternative channels given by the scattering
states defined by the two first eigenchannels as, $\ket{u_a}=a_1
\ket{\Phi_1} + a_2\ket{\Phi_2}$, with transmission $T_a=|a_1|^2 T_1 + |a_2|^2 T_2$
and similarly $\ket{u_b}$ with transmission $T_b$. Now the total transmission of
$\ket{u_c} = \ket{u_a} + \ket{u_b}$ will be $T_c = T_a + T_b + I$,
with the interference term $I=2 \, T_1 {\rm{Re}}(a_1 b_1^*) +
2 \, T_2{\rm{Re}}(a_2b_2^*)$. It will thus not simply be the sum of the
two transmissions, $T_c\neq T_a+T_b$.
Since the purpose of this paper is to
calculate the eigenchannels without solving for the complex band
structure in the leads, we will use the fact that the eigenchannels
also maximize the transmission through the device, i.e., the first
eigenchannel from the left contact maximize the transmission
probability over the space of  incoming states from the left, the
next channel maximizes the transmission while being orthogonal to
the first channel etc.

In the following, we will instead of flux-normalization make use of
{\em energy-normalization} $\braket{\Psi_n(E)}{\Psi_{m}(E')}=
\delta_{nm} \, \delta(E-E')$ with the trivial difference from
flux-normalization being a factor $\sqrt{\hbar}$.
Energy-normalization is advantageous when working with energy
resolved quantities since the natural continuous quantum number for
this normalization is the energy. In addition, we can interpret the
energy-normalized states as a density-of-states, i.e.,
$\vert\Psi_l(x)\vert^2=\vert \bra x \Psi_l\rangle\vert^2$ is the
projected (local) density-of-states at $x$.

\subsection{Preliminaries}
The one particle Hamiltonian
for the tight-binding scattering problem shown in \Figref{fig.sys}
can be written
\beq
H=H_0+V=H_L+H_D+H_R+V_{L}+V_{R} \, ,
\label{eq.hamiltonian}
\eeq
where the isolated leads $H_{L,R}$ and device $H_{D}$ are coupled
together by the interactions between leads and device
($V=V_{L}+V_R$) without any direct coupling between the leads.
Using projection operators onto the device $P_D$ and left/right
leads $P_{L,R}$ ($I=P_L+P_D+P_R$) we can also define
$\tau_{L,R}=P_D V_{L,R} P_{L,R}$ where $V_{L,R} =
\tau_{L,R}+\tau_{L,R}^\dagger$. The rest of this section provide a
short summary of the standard definitions of one particle Green's
functions, self-energies and the notation we will use throughout
the paper\cite{Datta,Paulsson.NEGF02}.

From the definition of the retarded Green's function (operator) for
the whole system we can find the expansion of the Green's function
in an eigenbasis, $H \ket{\Psi_m(E)}=E \ket{\Psi_m(E)}$,
\beqa
G(E)&=&\mybpar{E+i \delta-H}^{-1}= \nonumber \\
& = & \myint{}{} \mathrm{d}E' \,\mysum{m}{} \,
\frac{\ketbra{\Psi_m(E')}{\Psi_m(E')}}{E+i \delta -E'}
\label{eq.expansion} \, ,
\eeqa
where the infinitesimal imaginary part $\delta=0^+$ ensures that
the Green's function yield the retarded response of the system.
The device part of the Green's function can further be written
\beq
G_D=\mybpar{E+i \delta -H_D-\Sigma_L-\Sigma_R}^{-1}
\label{eq.gd} \, ,
\eeq
where we have introduced the self-energies,
$\Sigma_{L,R}=\tau_{L,R}^{\phantom{\dagger}}
g^{\phantom{\dagger}}_{L,R} \tau_{L,R}^\dagger$ given by the
Green's functions of the isolated leads $g_{L,R}=(E+i
\delta-H_{L,R})^{-1}$. In addition to the Green's function, the
spectral functions $A(E)=i(G-G^\dagger)$,
$a_{L,R}=i(g_{L,R}-g^\dagger_{L,R})$, and broadening
$\Gamma_{L,R}=i(\Sigma_{L,R}-\Sigma_{L,R}^\dagger)= \tau_{L,R}
a_{L,R} \tau^\dagger_{L,R}$ will be needed. The fact that these
matrices live on different subspaces will be used repeatedly,
e.g., $\Gamma_L=P_D \Gamma_L P_D$ etc.

For the scattering problem, see Fig.~\ref{fig.sys}, we know that the
time independent
discrete Schr\"odinger equation has a complete set of solutions.
These solutions can be divided into a continuous set of solutions
$\ket{\Psi_n(E)}$ (where there may be several solutions at any
given energy) and localized states $\ket{\Psi^{\mathrm{Loc}}_m}$
with energy $E^\mathrm{Loc}_m$. We use the energy as the
continuous quantum number together with a discrete quantum number
$n$, i.e., sub-bands. Since we are only interested in the transport
properties, we will from hereon ignore localized states
\footnote{Since we are interested in scattering
states around the Fermi-energy, we may choose an energy
arbitrarily close to the Fermi-energy where there are no localized
states.}.
%

In the following, we will describe a method to determine the
eigenchannel scattering states inside the device region using
the information contained in $G_D$, $\Gamma_L$, and $\Gamma_R$. The
spectral function, $A$, is a central quantity in the following
discussion. It can be obtained from the expansion of the retarded
Green's function in eigenfunctions to the Hamiltonian
(Eq.~\ref{eq.expansion}),
\beqa
A(E) & =& i\mybpar{G(E)-G^\dagger(E)} \nonumber\\
& = &  2 \pi \mysum{n}{} \ketbra{\Psi_n(E)}{\Psi_n(E)}
\label{eq.spectral} \, . \eeqa

\subsection{Scattering states from the leads}
We may choose to express the solutions to the Schr\"odinger equation as
solutions consisting of waves originating in the left or right lead.
These scattering states can be generated from the
spectral function as will be show here.
Decomposing the spectral function of the device $A_D=P_D A P_D$,
using Eq.~(\ref{eq.gd}), we find
\beqa
A_D&=&i (G_D-G_D^\dagger)=iG_D\mybpar{{G_D^{\dagger}}^{-1}-
       G_D^{-1}}G_D^\dagger \nonumber \\
&=&G_D \Gamma_L G_D^\dagger + G_D \Gamma_R G_D^\dagger
\label{eq.spectraldecomposition} \, ,
\eeqa
where we in the following wish to show that
$A_{L,R} =G_D \Gamma_{L,R} G_D^\dagger$ is generated by the
scattering states with incoming waves in the left (right) lead.

Viewing the coupling between device and leads as a perturbation,
we can start with a set of orthogonal and normalized
eigenfunctions, $\ket{{\widetilde{u}}_l}$, of the isolated left
lead (and similarly for the right) which are totally reflected
solution since they are solutions for the isolated semi-infinite
leads. From these states, the full solutions $\ket{{\Psi}_l}$ can
be generated,
\beq
\ket{{\Psi}_l}=G V_L \ket{{\widetilde{u}}_l}+
  \ket{{\widetilde{u}}_l}
\label{eq.normal} \, .
\eeq
The response given by the retarded Green's function only contains
waves traveling outwards from the device region. This solution to
the Schr\"odinger equation thus have the required property of
being incoming from the left lead. In addition, the solutions are
energy-normalized and orthogonal, see
Appendix~\ref{app.orthogonal}.

We can then express the device part of the spectral function
from the solutions generated by
Eq.~(\ref{eq.normal})
\beqa
A_{L}  &=&
2 \pi \mysum{l}{} P_D \ket{{\Psi}_l} \bra{{\Psi}_l} P_D \nonumber \\
&=&
2 \pi \mysum{l}{} P_D \mybpar{G V_L \ket{{\widetilde{u}}_l}+
   \ket{{\widetilde{u}}_l}} \nonumber \\ & &
   \mybpar{\bra{{\widetilde{u}}_l}+
   \bra{{\widetilde{u}}_l} V_L^\dagger G^\dagger}  P_D\nonumber \\
& = & P_D G V_L a_L V_L^\dagger G^\dagger P_D =
G_D^{\phantom{\dagger}} \Gamma_L G_D^\dagger \, , \label{eq.A_L}
\eeqa
where we have used Eq.~(\ref{eq.spectral}) for the whole system and
for the isolated lead.
Apart from re-deriving Eq.~(\ref{eq.spectraldecomposition}) we have
proven that the two parts of the device spectral function
$A_{L,R}=G_D \Gamma_{L,R} G_D^\dagger$
are built up of scattering states originating from the respective leads.
This immediately leads to the well known formula for the density
matrix in non-equilibrium (excluding localized states)
\beq
\rho = \frac{1}{2 \pi} \myint{-\infty}{\infty}
\left[f_L(E) A_L(E)+f_R(E) A_R(E) \right] \; \mathrm{d}E  \, ,
\eeq
where $f_{L,R}$ is the Fermi function of the leads.

\subsection{Current operator}
To find the eigenchannels of the system we will need the current
operator. The number of electrons in lead $R$ is described by the
projection operator $P_R$. The operator for current into $R$ is thus
determined the time derivative of $P_R$,
\beq \hat{J}_R=2 e
\dot{P}_R=\frac{i 2 e}{\hbar}[H,P_R]=\frac{i 2
e}{\hbar}(\tau_R^{\phantom{\dagger}}-\tau_R^\dagger)  \, , \eeq
where we have evaluated the commutator using the Hamiltonian,
Eq.~(\ref{eq.hamiltonian}), and included a factor of $2$ for spin.

The current into lead $R$ due to the scattering state with energy
$E$, $\ket{\Psi_l}$, originating from $L$ from the original
in-coming (standing) wave $\ket{\widetilde{u}_l}$, can be written
\beq j_{ll}= \brahket{\Psi_l}{\hat J}{\Psi_l} \, . \eeq To simplify
this expression, we extract the right lead part of the wavefunction
using the Lippmann-Schwinger Eq.~(\ref{eq.lippmann}), gives
\beq P_R
\ket{\Psi_l}=P_R G_0 V \ket{\Psi_l}+P_R \ket{\widetilde{u}_l} = g_R
\tau_R^\dagger \ket{\Psi_l} \, , \eeq where all quantities are
evaluated at energy $E$. The current carried by the scattering state
$\ket{\Psi_l}$ is then given by \beqa j_{ll}&=& \frac{i 2 e}{\hbar}
\brahket{\Psi_{l}}{\tau^{\phantom{\dagger}}_R-\tau^\dagger_R}{\Psi_l} \nonumber \\
&=& \frac{i 2 e}{\hbar} \left( \bra{\Psi_l}
 \tau_R
       g_R^{\phantom{\dagger}} \tau_R^\dagger
\ket{\Psi_l} - \bra{\Psi_l} \tau_R
     g_R^\dagger \tau_R^\dagger
 \ket{\Psi_l} \right) \nonumber \\
&=&  \frac{e}{\pi \hbar} \; 2 \pi  \bra{\Psi_l} \Gamma_R
\ket{\Psi_l} \; \label{eq.current1} \, . \eeqa Summing the current
over all the orthogonal, energy-normalized scattering states
originating in $L$ at the specific energy yield the net current or
total transmission, and thus the Landauer formula, \beqa
j&=&\mysum{l}{} J_{ll}= \frac{e}{\pi \hbar} \, 2 \pi \mysum{l}{}
\bra{\Psi_l} \Gamma_R  \ket{\Psi_l} \\
&=&\frac{e}{\pi \hbar} \, \mathrm{Tr}\mybpar{A_L \Gamma_R} =
\frac{e}{\pi \hbar} \,
   \mathrm{Tr}\mybpar{G_D \Gamma_L G_D^\dagger \Gamma_R } \, .
\nonumber
\eeqa

From Eq.~\ref{eq.current1}, we notice that the transmission
probability for any scattering state from the left $\ket{\Psi_l}$ is
given by $2 \pi \brahket{\Psi_l}{\Gamma_R}{\Psi_l}$. The
transmission probability matrix can therefore be written \beq
T_{l'l} = 2 \pi  \bra{\Psi_{l'}} \Gamma_R \ket{\Psi_l} \;
\label{eq.current2} \,, \eeq since we consider energy-normalized
(flux-normalized except for the factor of $\hbar$) scattering
states, $\ket{\Psi}$, for which the current is equivalent with the
transmission.

\subsection{Scattering states in the device region}

To find the eigenchannels from the left lead,
$\ket{\Phi_l}=\mysum{l'}{} ({\bf U}_L)_{ll'} \ket{\Psi_{l'}}$, we
need to diagonalize the transmission probability matrix. Since we do
not, in this paper, explicitly calculate the Bloch states in the
leads, we will diagonalize the transmission probability matrix in an
abstract basis formed by incoming waves from the left or right lead.
Another, equivalent, formulation is to find the eigenchannels by
maximizing the transmission at a given energy while keeping the
eigenchannels orthogonal (${\bf U}_L {\bf U}_L^\dagger={\bf 1}$)
\beq \mathrm{max} \; 2 \pi \brahket{\Phi_l}{\Gamma_R}{\Phi_l}
\label{eq.maximize}  \, . \eeq The main problem in finding the
eigenchannels is that we do not have access to the wavefunctions,
Green's functions or spectral functions of the entire system in a
typical calculation. We therefore have to find a method using only
properties from the device part of the system.

\begin{figure}[tb!]
\begin{center}
\includegraphics[width=0.95 \columnwidth,angle=0]{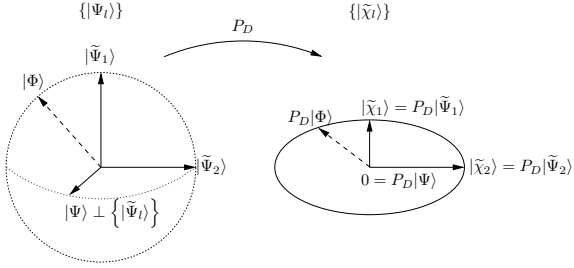}
\end{center}
\caption{Overview of the different states used in this section.
Note that the scattering states which are non-zero in the device subspace
are all spanned by the $\ket{\widetilde\Psi_l}$ states. We can
therefore find the eigenchannels by studying the
wavefunctions spanned by the
$\ket{\widetilde\chi_l}=P_D \ket{\widetilde\Psi_l}$.
\label{fig.space}}
\end{figure}

We will now show that
optimizing the current by varying the scattering state over space
spanned by the incoming states from the left lead,
$\left\{\ket{\Psi_l}\right\}$, is equivalent with varying over the
states $\ket{\widetilde\chi_l}$ defined on the device subspace,
see Fig.~\ref{fig.space}. We start by diagonalizing the device
part of the spectral function from the left lead
\beq
A_L(E) = \mysum{l}{}\ket{ \chi_l} \lambda_l \bra{\chi_l} =
2 \pi \mysum{l}{} \ket{\widetilde\chi_l} \bra{\widetilde\chi_l}   \, , \label{eq.chitilde}
\eeq
where the eigenvectors on the finite device space are
orthonormal $\braket{\chi_l}{\chi_{l'}}=\delta_{l,l'}$.
Each $\ket{\widetilde\chi_l}= \sqrt{\lambda_l/2 \pi} \ket{\chi_l}$
with non-zero eigenvalues ($\lambda_l \ne 0$) is
the device part of a specific state $\ket{\widetilde\Psi_l}$, i.e.,
$\ket{\widetilde\chi_l}=P_D \ket{\widetilde\Psi_l}$,
where the $\ket{\widetilde\Psi_l}$ states
are normalized and orthogonal linear combinations
of the scattering states $\ket{\Psi_l}$
\beqa
\ket{\widetilde\chi_l}&=&
\sqrt{\frac{\lambda_l}{2 \pi}} \ket{\chi_l}=
\frac{1}{\sqrt{2 \pi \lambda_l}} A_L(E)  \ket{\chi_l}= \nonumber \\
&=& P_D \sqrt{\frac{2 \pi}{\lambda_l}} \mysum{l'}{}\ket{\Psi_{l'}}
\braket{\Psi_{l'}}{\chi_l} \equiv P_D \ket{\widetilde\Psi_l} \, ,
\eeqa (using Eq.(\ref{eq.A_L})) where we have defined \beq
\ket{\widetilde\Psi_l}= \mysum{l'}{} \sqrt{\frac{2
\pi}{\lambda_{l'}}} \braket{\Psi_{l'}}{\chi_l} \ket{\Psi_{l'}}
\equiv\mysum{l'}{} W_{ll'} \ket{\Psi_{l'}}\,.
\eeq
This shows that
the states $\ket{\widetilde\Psi_l}$ are spanned by the incoming
scattering states from the left lead. We further wish to show that
these states are normalized in the same manner as the original
scattering states, i.e., we want to show that $W_{mn}$ is unitary
\beqa \mysum{n}{} W_{mn} W^*_{m'n} &=& \frac{2 \pi}{\sqrt{\lambda_m
\lambda_{m'}}} \mysum{n}{}
\braket{\chi_{m'}}{\Psi_n}\braket{\Psi_n}{\chi_m} \nonumber \\
&=&\frac{1}{\sqrt{\lambda_m \lambda_{m'}}} \bra{\chi_{m'}}
A_L \ket{\chi_m}\nonumber \\ &=&
\frac{\lambda_m}{\sqrt{\lambda_m \lambda_{m'}}} \braket{\chi_{m'}}{\chi_m}=
\delta_{mm'} \, ,
\eeqa
using Eqs. (\ref{eq.A_L}) and (\ref{eq.chitilde}).

In addition, any scattering state outside the
space spanned by $\ket{\widetilde\Psi_n}$
are orthogonal to the device subspace. This can
be seen from the fact that we can write
the spectral function on the device subspace as
\beqa
 A_L  & = & 2 \pi \mysum{l}{} P_D
\ketbra{\widetilde\Psi_l}{\widetilde\Psi_l}
P_D \hspace{1cm}\mbox{or as} \\
& = & 2 \pi \mysum{l}{} P_D \ketbra{\Psi_l}{\Psi_l} P_D \, . \eeqa
Comparing the two equations reveals that $P_D \ketbra{\Psi}{\Psi}
P_D$ must be zero for any $\ket{\Psi}$ which is orthogonal to the
space spanned by $\ket{\widetilde\Psi_l}$.

In this section we have shown that the wavefunctions
$\ket{\widetilde\chi_l}$ span the device part of any scattering
state generated from lead $L$, see Fig.~\ref{fig.space}. To find the
eigenchannels we can therefore maximize the current through the
device with respect to a linear combination of
$\ket{\widetilde\chi_l}$ instead of maximizing with respect to the
full scattering states $\ket{\Psi_l}$. This is equivalent with
diagonalizing the transmission matrix (Eq.~(\ref{eq.current2})) in
the basis formed by $\ket{\widetilde\chi_l}$.


\subsection{Finding the eigenchannels}


In general, the basis $\left\{\ket{e_i}\right\}$ used in a calculations 
is non-orthogonal with the overlap matrix defined by
$\mymat{S}{ij}=\braket{e_i}{e_j}$. Although the eigenchannels may be
calculated directly in this non-orthogonal basis, we will make use of
L\"ovdin ortogonalization to simplify the algebra. The ortogonalized 
matrices (denoted by a bar) are given by, 
${\bf \bar{\Gamma}}_R={\bf S}^{-1/2}{\bf \Gamma }_R{\bf S}^{-1/2}$
and ${\bf\bar{A}}_L={\bf S}^{1/2}{\bf A }_L{\bf S}^{1/2}$ etc. 
The eigenchannels obtained in the L\"ovdin orthogonalized basis 
can at the end of the calculation simply be transformed back into the 
non-orthogonal basis for further visualization or projection.

From the previous section we learned that it is enough to
diagonalize the transmission probability matrix
(Eq.~\ref{eq.current2}) $T_{l'l}=2 \pi \brahket{\widetilde\chi_{l'}}{
  \bar{\Gamma}_R}{\widetilde\chi_{l}}$
in the abstract basis  $\left\{\ket{\tilde \chi_{l}}\right\}$. To
transform $2 \pi {\bar{\Gamma}}_R$ into the basis
$\left\{\ket{\widetilde\chi_l}\right\}$, we first need to calculated
the eigenvectors of ${\bar{A}}_L=\bar{G}{\bar{\Gamma}}_L \bar{{G}}^\dagger$, 
\beq
\mysum{n}{}\left[{\bf{\bar{A}}}_L\right]_{mn} \mymat{U}{nl}=\lambda_l
\mymat{U}{ml} \, , \label{eq.practical1} 
\eeq 
where $\bf{U}$ is
unitary. We then obtain the transformation matrix to the
$\left\{\ket{\widetilde\chi_l}\right\}$ basis, 
\beq
\mymat{\widetilde{U}}{ml}=\sqrt{\frac{\lambda_l}{2 \pi}}
\mymat{U}{ml} \, , \label{eq.practical2} 
\eeq 
which gives the
explicit expression for the matrix we want to diagonalize 
\beq
T_{l'l}= 2\pi
\brahket{\widetilde\chi_{l'}}{\bar{\Gamma}_R}{\widetilde\chi_{l}}=
2\pi\left[\widetilde{\bf{U}}^\dagger {\bf{\bar{\Gamma}}}_R
\widetilde{\bf{U}}\right]_{l'l} \, , 
\eeq 
the eigenproblem is
therefore 
\beq 
\mysum{n}{}2 \pi
\left[\widetilde{\bf{U}}^\dagger{\bf{\bar{\Gamma}}}_R
\widetilde{\bf{U}}\right]_{mn} \mymat{c}{n\alpha}= T_\alpha
\mymat{c}{m\alpha} \, , \label{eq.practical3}
\eeq 
where the
eigenchannel vectors, $\bf{c}_{\alpha}$, are given in the basis 
described by the columns of $\widetilde{\bf{U}}$, and the
eigenvalues $T_\alpha$ are the transmission probabilities of the
individual eigenchannels, $\alpha$. Finally, transforming back to
the original non-orthogonal basis (from $\widetilde{\bf{U}}$-basis
to the L\"owdin-basis, and from L\"owdin basis to normal
non-orthogonal basis) we find that the eigenchannels on the device
subspace are given by, 
\beq 
P_D \ket{\Phi_\alpha} = \mysum{m,n}{} \left[
  {\bf{S}}^{-1/2} \widetilde{\bf{U}}\right]_{in}
\mymat{c}{n\alpha} \ket{{e}_i}\, . \label{eq.practical4} 
\eeq

The Eqs.~(\ref{eq.practical1})-(\ref{eq.practical4}) provide a
recipy for calculating the eigenchannels of a specific scattering
problem using only properties available in standard NEGF
calculations. In contrast to Ref.~\cite{Palacios.PRB06}, these eigenchannels
are well defined scattering states calculated without approximations on the
full device subspace.

It is interesting to note that the eigenchannels 
(Eq.~\ref{eq.practical4}) are eigenvectors to $G \Gamma_L G^\dagger
\Gamma_R$ (can be shown using the formalism presented above). This 
provides a simple method to obtain an idea about the eigenchannels. 
However, it is important to realize that the eigenchannel
wave-functions calculated from Eq.~\ref{eq.practical4} are energy
normalized, i.e., the amplitudes are well defined and can be compared 
between different eigenchannels. In contrast, the eigenvectors to 
$G \Gamma_L G^\dagger \Gamma_R$ may have any normalization and it is 
therefore not possible to compare amplitudes between different channels.
Moreover, the energy normalized scattering states Eq.~\ref{eq.practical4} 
yield amplitudes which correspond to local density of states and are
useful to plot, as will be shown in the next section.

\section{Eigenchannels for atomic- and molecular wires}
\label{sec.3}

To exemplify the method developend in the previous section
we will use three
different examples of atomic and molecular-wires connected to gold
electrodes. Using the
TranSIESTA\cite{BrMoOr.02.Density-functionalmethodnonequilibrium}
extension of the
SIESTA\cite{OrArSo.96.Self-consistentorder-Ndensity} density
functional theory (DFT) code, we have previously studied elastic and
inelastic transport properties for the systems under consideration
here, (i) atomic gold wires\cite{Frederiksen.PRB07}, the conjugated
organic molecules (ii) oligo-phenylene vinylene
(OPV)\cite{Paulsson.NANO06}, and (iii) oligo-phenylene ethynylene
(OPE)\cite{Paulsson.NANO06}. The TranSIESTA calculations on these
systems were performed using DFT in the generalized gradient
approximation using the PBE functional. The semi-infinite leads 
connecting the device region
was modeled using self-energies in the NEGF method. A more detailed
description of the calculational method may be found in
Ref.\cite{Frederiksen.PRB07}.

\begin{figure}[tbh!]
\begin{center}
\includegraphics[width=0.85 \columnwidth,angle=0]{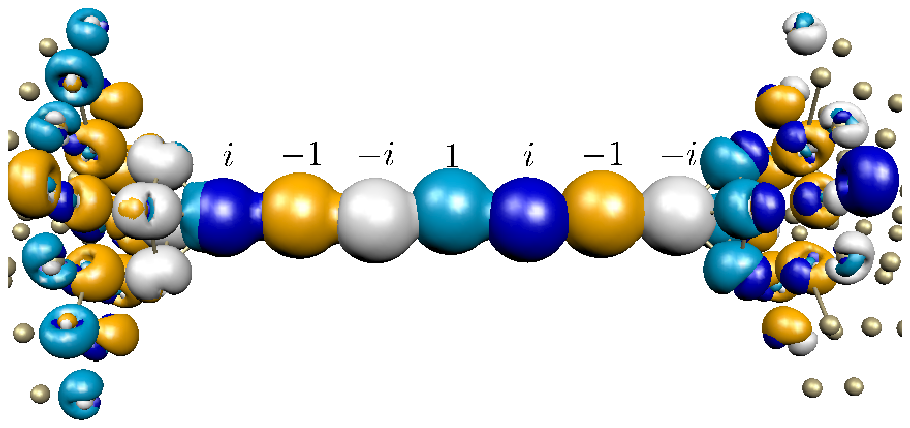}
\end{center}
\caption{(Color online) Left eigenchannel for a 7-atom Au atomic
chain connected by 4 atom pyramids to Au (100) surfaces. The complex
phase of the eigenchannel, idicated and shown in color, is similar that of a
Bloch wave at the Fermi level in the infinite atomic gold chain
(half-filled 6s-band).} \label{fig.auchain}
\end{figure}

Gold atomic wires has been realized and studied experimentally by
several different techniques. The low bias elastic and
inelastic (phonon scattering) transport is well characterized and understood,
see Ref.\cite{Frederiksen.PRB07} and references therein.
The first
eigenchannel (from left) at the Fermi-energy is shown in
Fig.~\ref{fig.auchain} for a 7-atom gold chain. Not surprisingly,
the majority of the transmission is carried by the first eigenchannel
($T_1=0.994$) with only a small transmission for the other eigenchannels
($< 10^{-5}$). In addition, it is clear that the current through the
wire is
carried by the 6-s electrons forming a half-filled one-dimensional band where
the sign of the wavefunction change by a factor of $i$ (right moving) along
the wire. The only difference with the corrsponding eigenchannel from the
right (not shown) is the phase factor which correcsponds to a left-moving
wave.

\begin{figure}[tbh!]
\begin{center}
\includegraphics[width=0.95 \columnwidth,angle=0]{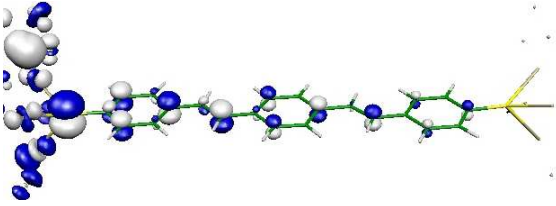}
\end{center}
\caption{(Color online) Left eigenchannel for an OPV molecule bound
by thiols to the hollow sites on  Au (111) surfaces. The colors
correspond to the two different signs of the almost real-valued
wavefunction. }\label{fig.OPV}
\end{figure}

For molecular wires, the experimental and theoretical understanding of
electron transport is less well understood. The calculated transmission
through the OPV
molecule shown in Fig.\ref{fig.OPV} is 0.037 and $100\%$ of the
transmission is carried through the first eigenchannel. Since the wave is
almost totally reflected, the imaginary part of the wave-function is too small
to be seen in the figure. In the calculation, the thiol bonds to the hollow 
site on the Au (111) surface and
clearly shows that the conjugation of the molecule continues through the
sulfur atom and that there is significant coupling to the gold leads.

\begin{figure}[tbh!]
\begin{center}
\includegraphics[width=0.95 \columnwidth,angle=0]{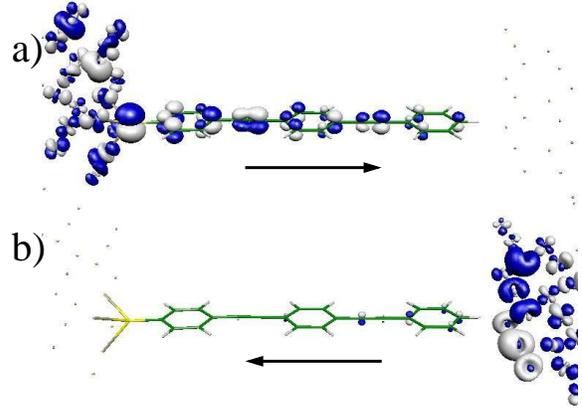}
\end{center}
\caption{(Color online) Left(a) and right(b) eigenchannel for an OPE
molecule strongly
  bound by a thiol to
  the left surface and weakly interacting with the right lead.
}\label{fig.OPE}
\end{figure}

To investigate an asymmetric case, we carried out calculations on an
OPE molecule bound by a thiol to the left lead, and with a tunneling
barrier (hydrogen termination) to the right hand lead, see
Fig.~\ref{fig.OPE}. The calculational details are the same as for
the OPE molecule in Ref.~\cite{Paulsson.NANO06}. Because of the
tunneling barrier, the transmission, $T_{tot}=0.0026$, is lower than
for the OPV molecule, and the left and right eigenchannels are
considerably different. This can easily be understood by the large
reflection at the right junction.

In the three examples described here, we find that the symmetry of
the eigenchannels can be intuitively understood from the band-structure 
of the corresponding infinite wires. For the gold-wire, 
the 5d-band is below the Fermi-energy which is situated approximately at
half-filling of the 6s-band.
The corresponding infinite wires (polymers) for the molecular wires have 
energy gaps at the Fermi-energy.
The eigenchannels therefore show the exponentially decaying solutions of the
$\pi$-electron state in the complex-bandstructure at the
Fermi-energy\cite{Sankey.PRB02}.

\section{Summary}
We have in this paper developed a method to calculate the scattering states
corresponding to elastic eigenchannels. The method is summarized in
Eqs.(\ref{eq.practical1})-(\ref{eq.practical4}) where the eigenchannels are
found from quantites normally available in transport calculations using the
NEGF technique. In addition, we show three
brief examples of elastic scattering states calculated for molecular and
atomic-wires
connected to three dimensional contacts. The eigenchannels for these systems
can be understood from the band-structure of the infinite wires providing an
intuitive understanding.

The eigenchannels are  useful to interpret elastic electron transport
through junctions. We belive that they will be especially useful to
investigate the effect of the contacts between device and leads, e.g.,
binding site of the thiol-bond on Au-surfaces. In addition, the method
gives a useful basis to understand the effects of phonon scattering on the
conductance and their propensity
rules\cite{Paulsson.PRB05,Paulsson.unpub07}.

\begin{acknowledgments}
The authors would like to thank
S. Datta, T. Frederiksen, C. Krag, and C. Rostgaard. for useful discussions.
This work, as part of the European Science Foundation EUROCORES
Programme SASMEC, was supported by funds from the SNF and the EC 6th
Framework Programme. Computational resources were provided by the
Danish Center for Scientific Computations (DCSC).
\end{acknowledgments}

\appendix

%
%

\section{Appendix: Orthogonality of scattering states}
\label{app.orthogonal} Viewing the Bloch states in the infinite,
periodic leads, $\ket{u_l}$, as a starting point for perturbation
theory, we can obtain the totally reflected solutions
$\ket{\widetilde{u_l}}$ of a semi-infinite lead. In this case the
perturbation is the removal of the coupling between the periodic
cells at the surface. Furthermore, the totally reflected states may
again be used as the starting point in a perturbation calculation to
obtain the full scattering states $\ket{\Psi_{l,r}}$. In this case
the perturbation is the device region and its coupling to the leads.
We will here show that the perturbation expansion gives solutions
that are orthogonal and normalized. To do this we focus on the
perturbation expansion of   $\ket{\Psi_{l}}$ from
$\ket{\widetilde{u_l}}$ and note that the same derivation may be
used to obtain the $\ket{\widetilde{u_l}}$ from $\ket{u_l}$.

Starting with a set of orthogonal and energy-normalized
eigenfunctions $\ket{{\widetilde{u}}_n(E)}$ of the isolated leads
($n\in l$, $r$) we generate the full scattering states
$\ket{{\Psi}_n(E)}$ \beq \ket{{\Psi}_n(E)}=G(E) V
\ket{{\widetilde{u}}_n(E)}+\ket{{\widetilde{u}}_n(E)}
\label{eq.normal2} \, , \eeq where $V=V_L+V_R$. The response given
by the retarded Green's function only contains waves traveling
outwards from the device region. To show that the solutions
generated in this way are normalized we use the Lippmann-Schwinger
equation \beq \ket{{\Psi}_n(E)}=G_0(E) V
\ket{{\Psi}_n(E)}+\ket{{\widetilde{u}}_n(E)} \label{eq.lippmann} \,
, \eeq where the unperturbed Green's function is
$G_0(E)=(E-H_L-H_R-H_D+i\delta)^{-1}$. Together with
Eq.~(\ref{eq.normal2}), we obtain
\begin{widetext}
\beqa
 \braket{{\Psi}_n(E)}{{\Psi}_{n'}(E')} & = &
  \braket{{\widetilde{u}}_n(E)}{{\Psi}_{n'}(E')} +
\brahket{{\widetilde{u}}_n(E)}{V^\dagger G^\dagger(E)}{{\Psi}_{n'}(E')} \\
  & = & \braket{{\widetilde{u}}_n(E)}{{\widetilde{u}}_{n'}(E')}+
  \brahket{{\widetilde{u}}_n(E)}{G_0(E') V^{\phantom{\dagger}}}{{\Psi}_{n'}(E')}
    +\brahket{{\widetilde{u}}_n(E)}{V^\dagger G^\dagger(E)}{{\Psi}_{n'}(E')} \\
 & = & \delta_{n,{n'}} \delta(E-E')+
  \brahket{{\widetilde{u}}_n(E)}{V^{\phantom{\dagger}}}{{\Psi}_{n'}(E')}
\mybpar{\frac{1}{E'-E+i\delta}+\frac{1}{E-E'-i\delta}}  \\
 & = & \delta_{n,{n'}} \delta(E-E')  \, ,
\eeqa
\end{widetext}
which shows that the final scattering states $\ket{\Psi_n}$
are orthogonal and normalized.

\end{document}